\documentclass{llncs}
\usepackage{makeidx}  

\def\etal{\emph{et al.}}

\usepackage{xcolor}
\usepackage{booktabs}       
\usepackage{siunitx}        
\usepackage{makecell}       
\usepackage{multirow}       

\usepackage[export]{adjustbox}
\usepackage{layouts}
\usepackage{graphicx}
\usepackage{lipsum}
\graphicspath{{./}{figures}}

\usepackage{subcaption}
\usepackage{caption}
\usepackage{subcaption}

\usepackage{amsmath, amssymb}
\usepackage{pifont}

\usepackage{hyperref}
\hypersetup{
    pdffitwindow=false,     
    pdfstartview={FitH},    
    colorlinks=true,       
    linkcolor=red,          
    citecolor=blue,        
    filecolor=cyan,         
    urlcolor=magenta        
}

\usepackage[utf8]{inputenc}
\usepackage[english]{babel}

\usepackage[
backend=biber,
style=ieee,
sorting=ynt
]{biblatex}
\usepackage{etoolbox} 
\makeatletter
\patchcmd{\ps@headings}{\rlap{\thepage}}{}{}{}
\patchcmd{\ps@headings}{\llap{\thepage}}{}{}{}
\makeatother
\begin{document}
%
%

%
\mainmatter              
%


\title{3D Guidewire Shape Reconstruction from Monoplane Fluoroscopic Images}

\author{Tudor Jianu\inst{1} \and Baoru Huang\inst{2} \and Pierre Berthet-Rayne\inst{3} \and Sebastiano Fichera\inst{4} \and Anh Nguyen\inst{1}}
\authorrunning{Ivar Ekeland et al.} 
%
\tocauthor{Ivar Ekeland, Roger Temam, Jeffrey Dean, David Grove,
Craig Chambers, Kim B. Bruce, and Elisa Bertino}

\institute{
Department of Computer Science, University of Liverpool, UK \and Imperial College London, UK \and 
Honorary Fellow, University of Liverpool, UK \and
Department of Mechanical, Materials \& Aerospace Engineering, University of Liverpool, UK 
}

\maketitle      

\begin{abstract}
Endovascular navigation, essential for diagnosing and treating endovascular diseases, predominantly hinges on fluoroscopic images due to the constraints in sensory feedback. Current shape reconstruction techniques for endovascular intervention often rely on either a priori information or specialized equipment, potentially subjecting patients to heightened radiation exposure. While deep learning holds potential, it typically demands extensive data. In this paper, we propose a new method to reconstruct the 3D guidewire by utilizing CathSim, a state-of-the-art endovascular simulator, and a 3D Fluoroscopy Guidewire Reconstruction Network (3D-FGRN). Our 3D-FGRN delivers results on par with conventional triangulation from simulated monoplane fluoroscopic images. Our experiments accentuate the efficiency of the proposed network, demonstrating it as a promising alternative to traditional methods.

\end{abstract}

\section{Introduction} \label{sec:intro}

Minimally invasive medical procedures have fundamentally transformed healthcare by offering alternatives that are less intrusive and lead to faster recovery times. A crucial aspect of their success lies in the accurate navigation and control of instruments, such as catheters and flexible manipulators. Traditionally, surgeons predominantly utilized 2D visualization techniques, with fluoroscopy being the most common. Yet, these traditional methods come with inherent challenges. Specifically, they limit depth perception, which complicates the visualization of intricate vascular structures~\cite{huang2021self, hoffmann2013reconstruction, huang2022self, hoffmann2015electrophysiology}. This limitation increases the risk of strenuous contact between instruments and arterial walls, thereby compromising patient safety and potentially reducing the procedure's effectiveness.

\begin{minipage}[h]{\textwidth}
    \begin{minipage}[t][][b]{.44\textwidth}
    \centering
    \includegraphics[width=\textwidth]{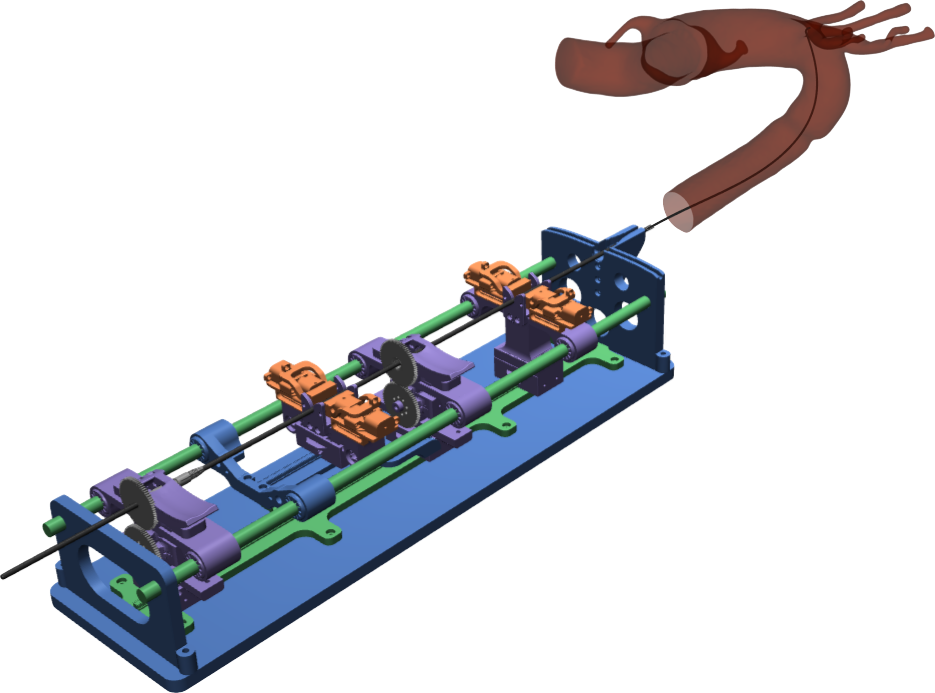}
    \vspace{-.3ex}
    \captionof{figure}{CathSim overview~\cite{jianu2022cathsim}}
    \label{fig:cathsim_intro}
    \end{minipage}
    \hspace{3ex}
    \begin{minipage}[t][][b]{.44\textwidth}
    \centering
    \includegraphics[width=\textwidth]{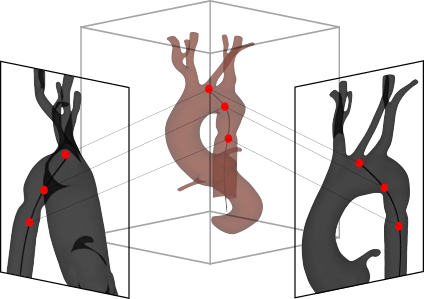}
    \vspace{-1.4ex}
    \captionof{figure}{Biplane fluoroscopic imaging reconstruction}
    \label{fig:triangulation}
    \end{minipage}
    \vspace{1ex}
\end{minipage}

To overcome these challenges, the accurate estimation of instrument shape is essential. This advancement not only enhances the surgeon's spatial understanding but also sets the stage for greater autonomy in instrument navigation. Such progress is readily observed in numerous robotic applications. When instruments are guided based on precise metrics, such as joint positions and velocities~\cite{kim2021observation} within the domain of robotic manipulation, the agent benefits of increased control over the task. This improvement in perception, by benefitting the efficiency, reduces the potential for surgeon fatigue and therefore human errors. 

In response to these challenges and recognizing the vast potential of improved visualization techniques, the scientific community has actively pursued advanced 3D reconstruction and localization methods. The shift from 2D to 3D visualization is designed to provide surgeons and autonomous systems with heightened spatial awareness. Such an improvement is anticipated to enhance instrument maneuverability, leading to more favorable patient outcomes~\cite{delmas2015three,nguyen2020end, vandini20133d}. This domain has seen a proliferation of strategies. These techniques generally rely on bi-planar scanners which are scarce and expose the patient to unnecessary radiation or include a priori information (such as kinematics), which depends on the accurate modeling of the guidewire~\cite{otake2014piecewise}. A third option offers the inclusion of fiber Bragg gratings (FBG) sensors paired with a single X-ray camera, which imposes further thickness restriction due to the inherent thickness of the fiber~\cite{taffoni2013optical}. While these methods offer good alternatives, challenges still persist in the reconstruction of guidewire in 3D during endovascular intervention. 

In our research, we introduce a significant leap forward in medical imaging with the 3D Fluoroscopy Guidewire Reconstruction Network (3D-FGRN). Leveraging the state-of-the-art endovascular simulator, CathSim~\cite{jianu2022cathsim}, and fusing it with deep learning techniques, our method distinguishes itself by its unique ability to perform 3D reconstructions straight from individual fluoroscopic images. This approach curtails the need for prior knowledge and reduces the dependence on specialized equipment, such as biplane fluoroscopic imaging systems and FBG sensors. As a result, there's a notable reduction in radiation exposure and alleviation of dimensionality constraints, all while maintaining a seamless surgical workflow. Moreover, the system's design permits integration into compact surgical tools, including guidewires.
\section{Related Work} \label{Sec:rw}

The domain of 3D reconstruction and localization for medical instruments has witnessed significant advancements, particularly in enhancing navigation and precision during interventions. This evolution is manifestly documented in numerous research studies.

Biplane fluoroscopic imaging systems offer simultaneous dual views to estimate surgical continuum robots' shape using the triangulation principle. Biplane-driven shape reconstruction methods primarily follow two strategies: 1) a bottom-up approach that detects the robot's centerline, as evidenced by the projective-invariant triangulation method proposed by Burgner~\etal~\cite{burgner2011toward} and improved versions by Wanger~\etal~\cite{wagner20164d} and Hoffman~\etal~\cite{hoffmann2015electrophysiology,hoffmann2012semi,hoffmann2013reconstruction}; and 2) a top-down approach that uses 3-D curves for approximation, as demonstrated by Delmas~\etal~\cite{delmas2015three}. While effective, these methods hinge heavily on biplane C-arm systems, which come with challenges such as high costs, significant radiation doses, and operational workspace constraints during interventions. Consequently, there's a pivot towards monoplane C-arm-based shape-reconstruction techniques, supplemented with additional kinematic model data, to address these limitations.

Monoplane fluoroscopic imaging systems have been instrumental in developing shape-sensing methods for surgical robots. Lobaton~\etal devised a technique for concentric tube robots in bronchoscopy surgery that blended deformable surface parameterization with data from selected fluoroscopic images, optimizing C-arm positioning using probabilistic priors~\cite{lobaton2013continuous}. Although precise, it was limited to simulated data and offline deformation modeling. Vandini~\etal, using a similar monoplane approach, reconstructed a Hansen artisan robotic catheter's shape by leveraging offline appearance priors and extracting 2-D centerlines~\cite{vandini20133d}. However, both techniques necessitated C-arm positioning adjustments, presenting workflow challenges. Addressing this, Vandini~\etal introduced an automated method for transnasal surgery robots, integrating visual information from monoplane fluoroscopy with kinematic models, eliminating the need for C-arm repositioning~\cite{vandini2015vision}. Otake~\etal~\cite{otake2014piecewise} and Papalazarou~\etal~\cite{papalazarou20123d} further expanded on these methods, integrating intraoperative monoplane fluoroscopy with piecewise-rigid 2D/3-D registration and nonrigid structure from motion, respectively. Yet, some methods required multiple monoplane views with minimal separation for accurate 3-D pose determination. Despite these advancements, monoplane fluoroscopic methods often disrupt clinical workflows, expose patients to significant radiation, and demand high computational resources, underscoring the need for further refinement in future clinical applications.

While both biplane and monoplane strategies have their benefits, they confront various challenges. Biplane systems are encumbered by high costs, increased radiation doses, and workspace limitations, whereas monoplane methods often disrupt clinical workflows, intensify radiation concerns, and impose computational strains. In contrast, deep learning can provide an enhanced understanding of monoplane images, thereby predicting the robot shape with remarkable accuracy. However, this solution is not without its own set of challenges, particularly the substantial data requirements.

\section{Method}

\subsection{Preliminary}

\paragraph{CathSim Simulator.} Our research leverages CathSim, an advanced endovascular simulator, to underpin the development of our shape prediction models. CathSim comprises four primary components: \textit{i)} the follower endovascular robot, \textit{ii)} the aortic arch phantom, \textit{iii)} the guidewire model, and \textit{iv)} the blood simulation. Our preference for CathSim stems from its notable efficiency in sample generation and its compatibility with Reinforcement Learning algorithms. This compatibility allows for the extraction of a wealth of training samples. A visual representation of the entire simulation environment is provided in Fig.~\ref{fig:cathsim_intro}.

\paragraph{Sampling Mechanism.} To address our data collection needs, we use a policy \( \pi \) trained in our specific environment. This network features a multi-modal feature extractor. Within this setup, the image input \( I \), in conjunction with the segmented guidewire \( S \), is integrated with intrinsic data, namely the joint positions \( J_p \) and joint velocities \( J_v \). The policy's training is grounded in the soft actor-critic (SAC) approach~\cite{haarnoja2018soft}.

\subsection{Shape Reconstruction}\label{sec:shape-reconstruction}

To achieve precise reconstruction of the guidewire's shape, we adopt a two-fold approach: first, utilizing image processing techniques, and then, implementing a triangulation process with two orthogonal cameras. We begin by obtaining segmented images of the guidewire from both camera perspectives. From these, we extract the guidewire's backbone or its central points. By applying a predetermined spacing parameter, we ensure that points are uniformly sampled along the guidewire's skeleton, thus facilitating the reconstruction.

The extraction of backbone points necessitates the identification of the guidewire skeleton within the segmented image. The skeleton is obtained via morphological thinning—a technique deployed to erode the boundaries of objects in binary images until a representation is reduced to a single-pixel width, thereby preserving the guidewire’s topology~\cite{guo1989parallel}. 

The extraction process is initiated from an endpoint of the skeleton—a point with only one neighbor—and advances through the skeleton, selecting points at intervals as determined by the predetermined spacing parameter. This approach enables the extraction of the guidewire's backbone points, preserving the topological structure and streamlining subsequent triangulation processes.

The 2D points extracted from both the top and side views are triangulated to determine the 3D representation of the guidewire. Given two cameras with projection matrices \(P_1\) and \(P_2\), and their corresponding homogeneous image points \(x_1\) and \(x_2\), the aim is to triangulate these points to find the homogeneous 3D coordinate, \(X\).

For a given 3D homogeneous point, represented as \(X = [X, Y, Z, W]^T\), and the projection matrix \(P\):

\begin{equation}
P = \begin{bmatrix}
f_x & s & c_x \\
0 & f_y & c_y \\
0 & 0 & 1 \\
\end{bmatrix}
\begin{bmatrix}
r_{11} & r_{12} & r_{13} & t_1 \\
r_{21} & r_{22} & r_{23} & t_2 \\
r_{31} & r_{32} & r_{33} & t_3 \\
\end{bmatrix}
\end{equation}

The projection of \(X\) onto the 2D homogeneous point \(x = [u, v, w]^T\) is given by:

\begin{equation}
    x = P \times X
\end{equation}

Given the projections from both the first and second cameras, we can construct a system of linear equations as:

\begin{equation}
    \begin{aligned}
        x_1 \times (P_1 X) = 0  \\
        x_2 \times (P_2 X) = 0 
    \end{aligned}
\end{equation}

\begin{figure}[t]
    \centering
    \hfill
    \includegraphics[width=0.9\linewidth,height=0.7\linewidth]{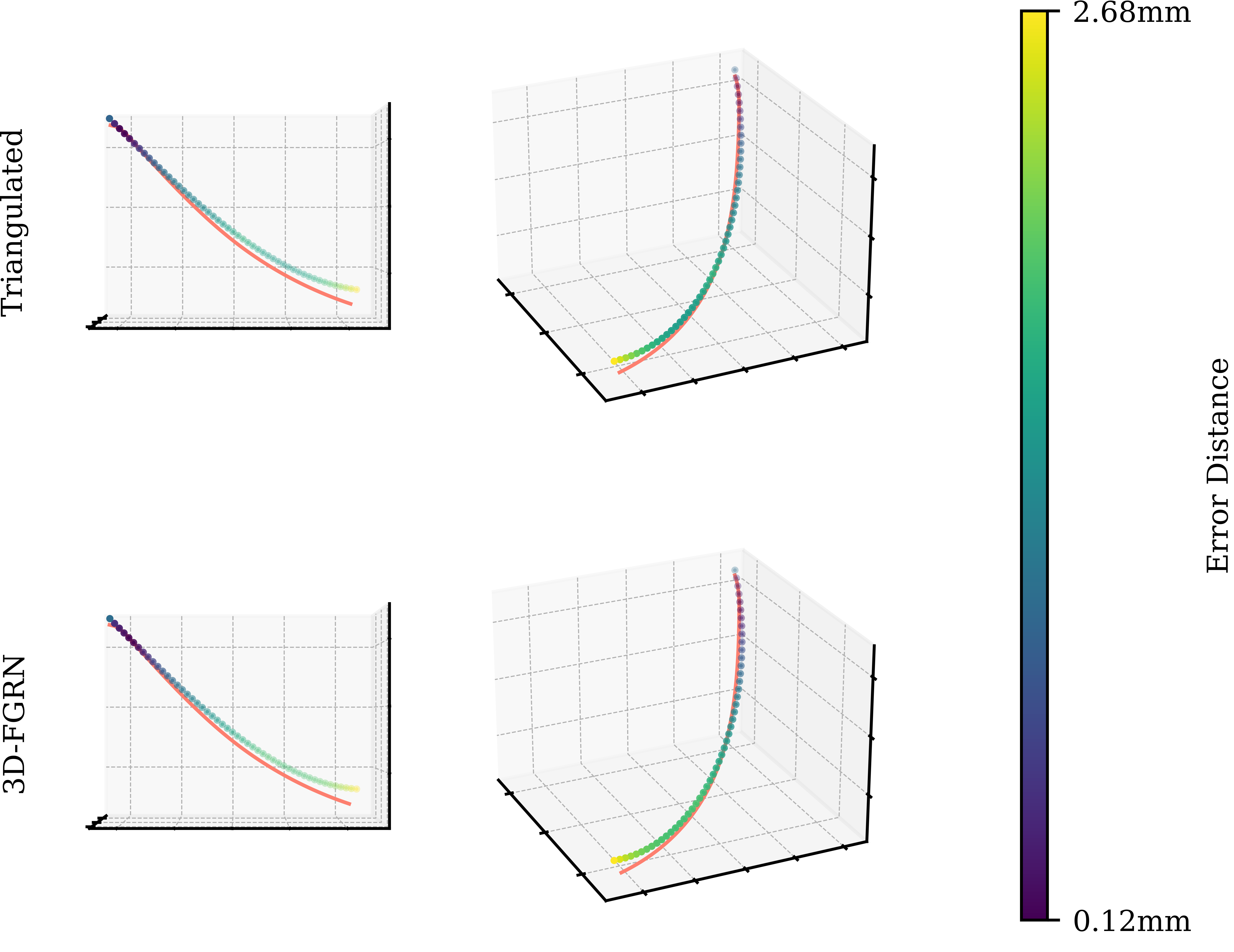}
    \caption{Comparison of the triangulated shape (upper row) and the 3D-FGRN's prediction (bottom row). Points are generated using univariate spline interpolation~\cite{dierckx1995curve}, sampled at intervals of \(2\unit{mm}\). These points are color-coded based on their distance from the corresponding ground-truth point. As observed, the two shapes exhibit a strong resemblance.}
    \label{fig:3D-reconstruction}
\end{figure}

Elaborating on these equations results in a system of the form \(AX = 0\), wherein \(A\) is a \(4 \times 4\) matrix composed of rows from the projection matrices \(P_1\) and \(P_2\) and the image coordinates \(x_1\) and \(x_2\). By employing the Singular Value Decomposition (SVD)~\cite{klema1980singular} on \(A\), the desired \(X\) minimizing \(||AX||\) subject to \(||X|| = 1\) is determined. The solution corresponds to the eigenvector associated with \(A\)'s smallest singular value, furnishing the homogeneous coordinates for the triangulated 3D point.

Following the triangulation, we apply a univariate spline interpolation to the 3D points, as described by Dierckx~\cite{dierckx1995curve}. This results in a smooth curve in 3D space. The representation ensures the possibility of sampling points at equal arc lengths, facilitating a direct comparison with the actual guidewire representation. The triangulation process can be viewed in Fig.~\ref{fig:triangulation}.

\subsection{Shape Prediction Network}\label{sec:shape-prediction}

After deriving the reconstruction of the guidewire shape, 3D-FGRN is trained to approximate the instrument shapes from 2D images. 3D-FGRN is composed of two main parts: a convolutional feature extraction module and a linear classification module. The convolutional module consists of a sequence of convolutional layers designed to capture spatial hierarchies within the input data. Each convolutional layer is followed by a Rectified Linear Unit (ReLU)~\cite{agarap2018deep} activation function, which introduces non-linear transformations essential for intricate pattern delineation. Subsequently, max-pooling layers~\cite{nagi2011max} are interspersed after each activation, aiming to reduce the spatial dimensionality and enhance the representational potency of the extracted feature maps. The linear classification module, following the convolutional section, employs a series of fully connected layers that forge high-level abstractions from the convolutional layers. Within this module, a ReLU activation function is again utilized to ensure non-linearity, and a dropout layer~\cite{srivastava2014dropout} is introduced as a regularization technique, designed to mitigate the risk of overfitting by randomly nullifying certain neurons during the training phase. Collectively, this architecture is poised to effectively learn and make predictions based on shape-related features from three-dimensional data.

The training dataset is composed of image observations of shape \(80 \times 80\) paired with the derived 3D positions, \(p\), of the guidewire bodies, enabling the model to learn the mapping from 2D images to the reconstructed 3D shapes. The CNN is optimized using the NAdam~\cite{dozat2016incorporating} optimizer, with a refined loss function that combines the Huber Loss and a regularization term to ensure both the accuracy of the predictions and the preservation of the geometric structure of the guidewire.

The Huber Loss, \(\mathcal{L}_{\text{Huber}}\), is defined as:

\begin{equation}
    \mathcal{L}_{\text{Huber}}(y, \hat{y}) = \begin{cases} 
                                    \frac{1}{2}(y - \hat{y})^{2} & \quad \text{if } \left | y - \hat{y} \right | < 1 \\
                                    \left| y - \hat{y} \right| - \frac{1}{2} & \quad \text{otherwise}
                                \end{cases}
\end{equation}

Here, \(y\) represents the ground truth, the derived 3D positions of the guidewire bodies, and \(\hat{y}\) represents the predicted positions by the model.

To maintain the geometric integrity of the guidewire, a regularization term, \(\mathcal{L}_{\text{reg}}\), is introduced, which penalizes the deviation of the interbody spacing of \(s=0.002\unit{m}\):

\begin{equation}
    \mathcal{L}_{\text{reg}} = \frac{1}{n-1} \sum_{i=1}^{n-1} \left| \lVert \hat{y}_{i+1} - \hat{y}_i \rVert_2 - s \right|
\end{equation}

The final loss function, \(\mathcal{L}\), is a weighted sum of the Huber Loss and the regularization term, ensuring both accurate predictions and preservation of the guidewire’s geometric structure:

\begin{equation}
\mathcal{L}= \alpha \mathcal{L}_{\text{Huber}}(y, \hat{y}) + \beta \mathcal{L}_{\text{reg}} 
\end{equation}

Similarly to the 3D reconstruction, we apply a univariate spline interpolation~\cite{dierckx1995curve} to the 3D points, thus getting a smooth representation of the guidewire.

\section{Experiments} \label{Sec:exp}

Our experiments commence with the use of a trained policy \( \pi \) in a simulated Type-I Aortic Arch environment. The samples produced from this procedure subsequently form a dataset dedicated to the shape-prediction task, yielding a dataset size of \(N=10,000\).

\subsection{Qualitative Analysis}

\begin{figure}[t]
    \centering
    \includegraphics{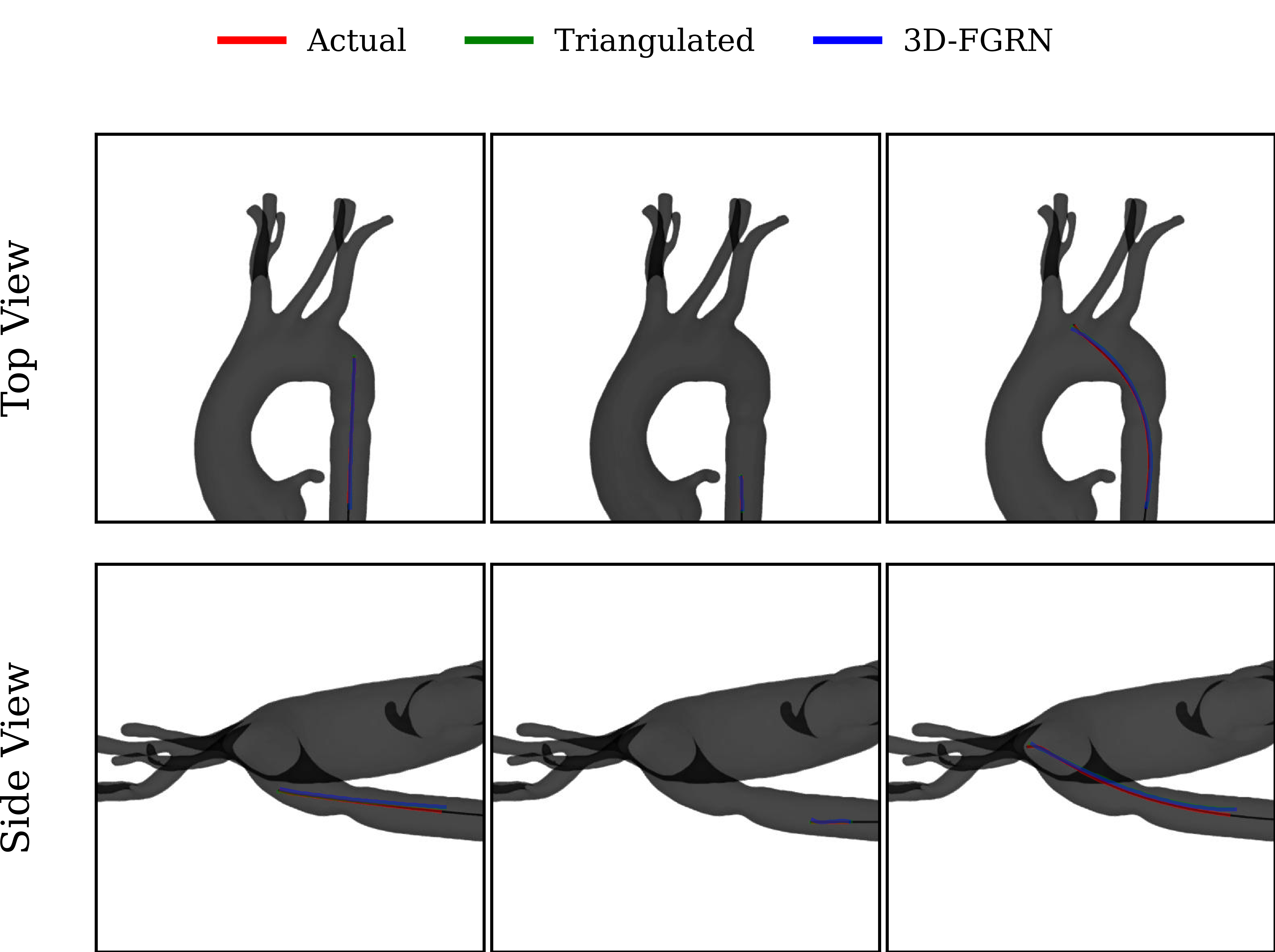}
    \caption{Comparison showcasing the actual guidewire shape, the triangulated configuration, and the 3D-FGRN's prediction. The visual juxtaposition emphasizes the striking resemblance among the three representations.}
    \label{fig:actual_vs_pred_2d}
\end{figure}

By employing the method detailed in Section~\ref{sec:shape-reconstruction}, we extract \(n_t\) points from the top image and \(n_s\) points from the side image. These extracted points are used in deriving the 3D coordinates that represent the guidewire's shape. Concurrently, the specific architecture of the guidewire is inferred. The 3D-FGRN is trained using these triangulated points, with the results showcased in Fig.~\ref{fig:3D-reconstruction}. The predicted shape aligns closely with the actual configuration of the guidewire in both instances. Notably, the proximal end manifests a more substantial error relative to the nominal error seen at the distal end. Discrepancies from the authentic guidewire shape span from a mere \(0.12\unit{mm}\) at the distal end to a noticeable \(2.68\unit{mm}\) at the proximal end. Impressively, 3D-FGRN evidences its capability to accurately predict the guidewire's shape using only a singular plane image. Subsequently, the 3D points are reprojected onto the original images, as illustrated in Fig.~\ref{fig:actual_vs_pred_2d}. Mirroring the 3D representation, there exists only a slight difference between the triangulated shape and the 3D-FGRN prediction.

\begin{figure}[t]
    \centering
    \includegraphics{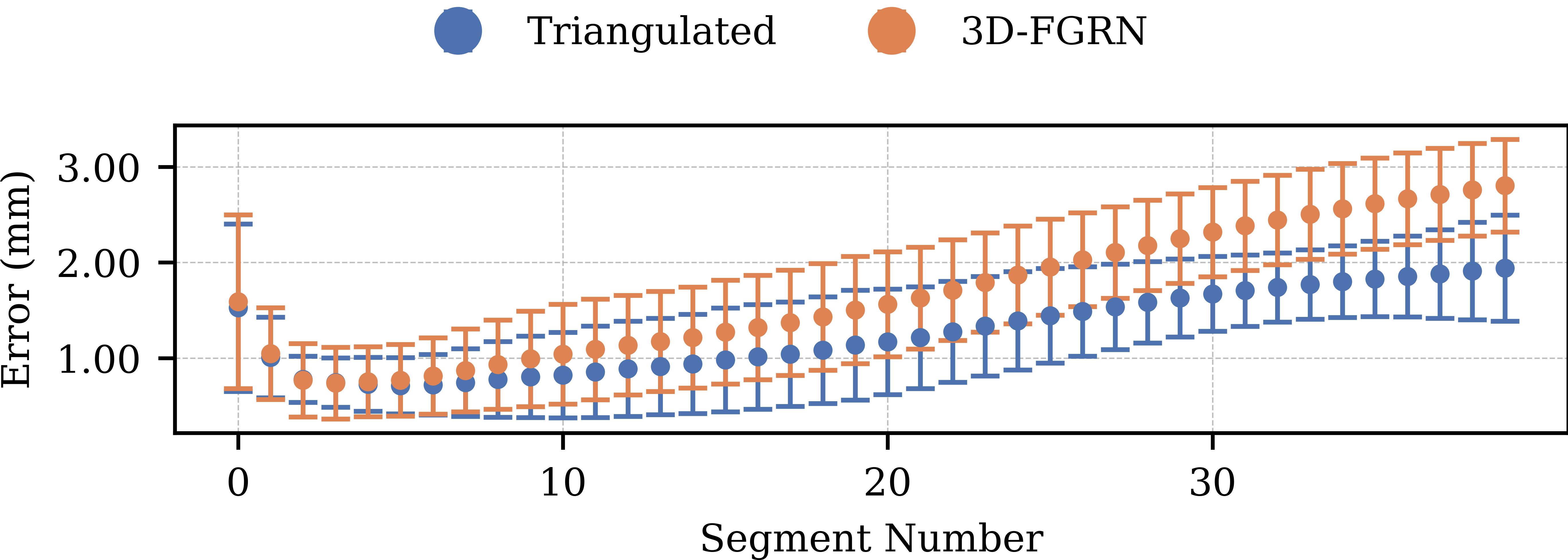}
    \caption{Comparison of matching points along the guidewire, from the distal end (segment~\(0\), representing the guidewire tip) progressing towards the proximal end. The error trends suggest a propagation towards the proximal end, with a notably smaller error at the distal segments. The pattern further underscores the congruence between the triangulated shape and the 3D-FGRN's prediction.}
    \label{fig:error-plot}
\end{figure}

Expanding on our previous experiments, we assessed the mean error between equidistant points along the guidewire. Given that our sampling initiated from the distal end — the guidewire's tip — and proceeded toward the proximal end, the segments are uniform, enabling a direct comparison. The disparity in errors between the triangulated method and the 3D-FGRN is showcased in Fig.~\ref{fig:error-plot}. This representation underscores our earlier findings: errors tend to increase from the distal to the proximal end. It's worth noting that the tip of the guidewire also presents a heightened error, possibly attributed to its distinct curvature. Importantly, the patterns of error are closely aligned, indicating the network's proficiency in accurately determining shape from a limited observation scope. However, we note a higher error of the prediction compared to the triangulation, especially at the distal end.

\begin{table}[th]
    \setlength{\tabcolsep}{12pt}
    \centering
    \caption{Comparison of shape deviation.}
    \vspace{0.2cm}
    \begin{tabular}{l r r r r}
        \toprule
        & \thead{MaxED~\(\downarrow\)} & \thead{METE~\(\downarrow\)} & \thead{MERS~\(\downarrow\)} \\
        \midrule
        Reconstruction & 2.880 \(\pm\) 0.640 & 1.527 \(\pm\) 0.877 & 0.001 \(\pm\) 0.000 \\
        3D-FGRN        & 3.479 \(\pm\) 0.819 & 1.589 \(\pm\) 0.909 & 0.001 \(\pm\) 0.001 \\
        \bottomrule
    \end{tabular}
    \label{tab:shape-accuracy}
\end{table}

\subsection{Quantitative Analysis}

To provide an objective comparison between guidewire shapes obtained from both the reconstruction method and the 3D-FGRN network, we utilized three metrics: i) MaxED, which signifies the maximum Euclidean distance; ii) METE, indicating the mean error in tip tracking; and iii) MERS, reflecting the mean error related to the robot's shape. Lower values for METE, MERS, and MaxED generally represent reduced discrepancies between the reconstructed and the actual shapes. A side-by-side assessment between the conventional reconstruction approach and the 3D-FGRN prediction can be found in Table~\ref{tab:shape-accuracy}. The empirical findings demonstrate a notable consistency between METE and MERS values across both methods. However, a distinct difference is evident in the MaxED metric, where the 3D-FGRN approach yields a value of \(3.479 \pm 0.819\), in contrast to the \(2.880 \pm 0.640\) secured by the reconstruction. This variation underscores that while the network exhibits commendable efficacy overall, there remains room for further optimization in subsequent versions.
\section{Conclusions and Future Work}\label{sec:conclusion}

This study emphasizes the development of a network adept at shape reconstruction, especially when working within a limited observational domain, such as 2D images. Despite certain constraints, the 3D-FGRN has demonstrated an impressive capacity to emulate the outcomes of traditional triangulation techniques. As we continue to refine our triangulation approaches, we expect the accuracy and reliability of 3D-FGRN to progress in tandem. In light of our findings, our next step involves the creation of an extensive dataset from a biplanar scanner. Our intention is to contribute to the scientific community by introducing a benchmark dataset, designed to evaluate and improve shape reconstruction techniques whilst testing our results in the real domain. Through this initiative, we aim to provide tools and insights that will be pivotal in enhancing surgical accuracy and patient care, ultimately steering towards greater surgical autonomy.

\medskip
\printbibliography
\pagebreak

\end{document}